\documentclass[amssymb,fleqn,aps,twocolumn,showpacs,nofootinbib,superscriptaddress]{revtex4}
\usepackage[]{graphicx}
\usepackage[]{amsmath}
\usepackage{setspace}
\usepackage{hyperref}
 \usepackage{setspace}

\def\Tr{\mathrm{Tr}}

\def\log{\mathrm{log}}

\def\cN{\mathcal{N}}

\def\cZ{\mathcal{Z}}

\def\eq#1{Eq.~\eqref{eq:#1}}

\def\({\left(}
\def\){\right)}

\begin{document}

\title{Coarse grained belief propagation for simulation of interacting \\ quantum systems at all temperatures}
\author{Ersen Bilgin}
\email{ersen@caltech.edu}
\affiliation{Institute for Quantum Information, California Institute of Technology, Pasadena, CA 91125}
\author{David Poulin}
\email{David.Poulin@USherbrooke.ca}
\affiliation{D\'epartement de Physique, Universit\'e de Sherbrooke, Sherbrooke, Qu\'ebec, Canada}

\date{\today}

\begin{abstract}
We continue our numerical study of quantum belief propagation initiated in \cite{PB08a}. We demonstrate how the method can be expressed in terms of an effective thermal potential that materializes when the system presents quantum correlations, but is insensitive to classical correlations. The thermal potential provides an efficient means to assess the precision of belief propagation on graphs with no loops. We illustrate these concepts using the one-dimensional quantum Ising model and compare our results with exact solutions. We also use the method to study the transverse field quantum Ising spin glass for which we obtain a phase diagram that is largely in agreement with the one obtained in \cite{LSS07a} using a different approach.  Finally, we introduce the coarse grained belief propagation (CGBP) algorithm to improve belief propagation at low temperatures.  This method combines the reliability of belief propagation at high temperatures with the ability of entanglement renormalization to efficiently describe low energy subspaces of quantum systems with local interactions.  With CGBP, thermodynamic properties of quantum systems can be calculated with a high degree of accuracy at all temperatures.  
 \end{abstract}


\maketitle

\medskip 
\section{Introduction} 
The most interesting phenomena in condensed matter physics occur when a large number of quantum particles are put into interaction. However, even the most simplified models of these interactions rarely admit analytical solutions. For this reason, numerical methods must be called into play. As direct numerical simulations require resources that scale exponentially with the number of particles, approximation schemes are needed to understand these systems at their thermodynamic limit. 

Starting with White's density matrix renormalization group (DMRG) method  \cite{Whi92a}, families of states have been introduced that accurately describe the low energy sector of locally interacting quantum systems using few parameters \cite{AKLT88a,V03a,VC04a}. In particular, matrix product states (MPS) approximate ground states of one-dimensional gapped Hamiltonians \cite{VC06a,Has07a}, and projected entangled pair states (PEPS) are accurate in higher dimensions at finite temperature \cite{Has06a,H07a}.

Even though these states require few parameters, finding the right set of parameters for a given system remains a formidable task \cite{SCV08a}. Only in special cases, such as simulating imaginary time evolution within a variational set of states \cite{Vid06a}, or using iterative minimization procedures~\cite{VC04a}, have successful heuristics been devised that solve this problem.

In this article, we introduce {\em coarse grained belief propagation} (CGBP), which combines the strength of two such heuristics. The first method, devised by Vidal, is a refinement of real space renormalization \cite{Wil75a} called {\em entanglement renormalization} (ER) \cite{Vid05a,EV08a}. The strength of this method comes from the realization that entanglement is organized on different length-scales in the ground state of many systems of interest, including critical systems \cite{Vid05a,V08a,CDR08a} and systems with topological order \cite{AV08a,KRV09a}.  As a consequence of this organization, entanglement can be efficiently removed from the state by a sequence of local coarse-graining transformations. This leads to a very efficient scheme for finding low-energy states of local Hamiltonians, making ER a very effective method for low temperatures.

The second method is belief propagation (BP), which is a well studied and widely used method to solve inference problems involving a large number of correlated random variables (see e.g. \cite{P88a,AM00a,YFW02a}). This algorithm is exact on trees where it essentially reduces to a transfer matrix solution. On more general graphs, it can be described as performing a constrained minimization of the Bethe free energy of the system \cite{YFW02a}.  Therefore, it  is often a very good approximation on graphs containing no small loops. On graphs with small loops, generalizations of BP can also provide reliable approximations ~\cite{YFW02a,WJW03a}.

Belief propagation was recently generalized to quantum theory by us and others \cite{Has07b,LP08a,PB08a,LSS07a}. The gist of the method consists of solving the system exactly on a small cluster and using this solution to compute effective thermal Hamiltonians on the neighboring clusters. This procedure is repeated iteratively until it produces a correction to the bare Hamiltonian that accurately mimics a system of infinite size. The method is very accurate at high temperatures and on trees or graphs with only large loops \cite{PB08a}. However, BP becomes unreliable at low temperatures because the clusters must be larger or equal to the range of the effective Hamiltonian, which grows like the inverse of the temperature. 

Coarse grained belief propagation combines the features of BP and ER. Starting at high temperatures where plain BP is accurate, temperature is lowered until coarse-graining the lattice by eliminating its shortest length-scale degrees of freedom becomes favorable. Coarse graining discards some high energy states, which results in a systematic error in the thermal state. On the other hand, it increases the effective size of the clusters, making BP more accurate. The coarse-graining procedure is continued as temperature is lowered to zero where plain ER is accurate. An estimate of the error caused by BP can be used to determine the temperatures at which each coarse-graining procedure should be performed. 

In Sec. \ref{sec:BP}, we review the BP method and describe how to estimate its accuracy. Based on this technique, Sec. \ref{sec:Cayley} presents results obtained for the transverse field quantum Ising spin glass studied in \cite{LSS07a}.  Finally, in Sec \ref{sec:CGBP} we introduce the CGBP algorithm, and benchmark it using an infinite quantum Ising chain.

\section{Belief propagation}
\label{sec:BP}

\subsection{Classical setting}
\label{classicalBPsection}

Consider a system composed of $N$ classical spins on a chain with Hamiltonian $H(x_1,\ldots,x_N) =\sum_{\langle i,j\rangle} h(x_j,x_j)$, where $\langle i,j\rangle$ denotes neighboring sites. The partition function for this system is defined as
\begin{align}
\cZ(\beta) &= \sum_{x_1,\ldots, x_N} e^{ -\beta H(x_1,\ldots, x_N)} \\
& = \sum_{x_1,\ldots, x_N} e^{-\beta h_{N,N-1}} \ldots e^{-\beta h_{3,2}} e^{-\beta h_{2,1}}\label{eq:Zclass}
\end{align}
where we use the shorthand $h_{i,j} = h(x_i,x_j)$. A brute force calculation of this quantity requires summing over an exponential (in $N$) number of terms. However, taking advantage of the local structure of the Hamiltonian, the sum can be rearranged as follows:
$$
\sum_{x_N}e^{-\beta h_{N-1,N}} \left(\ldots \sum_{x_2} \left( e^{-\beta h_{2,3}} \sum_{x_1} e^{-\beta h_{1,2}} \right)\ldots \right).
$$

Now, the sums can be performed sequentially with each sum only involving a small number of terms\footnote{The procedure we are describing for a chain is usually referred to as the transfer matrix method in statistical physics.}, allowing the computation of $\cZ$ in a time proportional to $N$. This leads to an iterative rule where messages $m_{i\rightarrow j}$ are exchanged along the edge of the chain connecting two sites $i,j$ with the update rule $m_{i \rightarrow i+1}(x_{i+1}) = \sum_{x_{i}} e^{-\beta h(x_{i},x_{i+1})} m_{i-1\rightarrow i}(x_i)$. With the initialization  $m_{0\rightarrow 1}(x_1) = 1$, we see that $\cZ(\beta) = \sum_{x_N} m_{N-1\rightarrow N}(x_N)$.  

This procedure can be generalized to arbitrary graphs by defining the update rule for the message $m_{i\rightarrow j}(x_j)$, passed from site $i$ to a neighboring site $j$, to be
\begin{equation}
m_{i\rightarrow j}(x_j) = \sum_{x_i} e^{-\beta h(x_i,x_j)} \prod_{k \in \cN(i)\backslash j} m_{k\rightarrow i}(x_i)
\label{eq:MP}
\end{equation}
where $\cN(i)\backslash j$ denotes the set of neighbors of site $i$ other than $j$. On a tree, these messages will converge to their final value after a time equal to the tree's diameter. The one- and two-body beliefs $b_j(x_j) = \frac{1}{\cZ(\beta)} \prod_{k \in \cN(j)} m_{k\rightarrow j} (x_j)$ and $b_{i,j}(x_i,x_j) = \frac{1}{\cZ(\beta)} \prod_{k \in \cN(i)} m_{k\rightarrow i} (x_i) \prod_{k' \in \cN(j)} m_{k'\rightarrow j} (x_j) e^{-\beta h(x_i,x_j)}$ are equal to the reduced one- and two-body distribution respectively and the partition function can be evaluated from any one of their normalization. When the underlying graph contains loops, BP is no longer exact but often provides accurate approximation to the true marginal states and partition function.

\subsection{Quantum setting}
\label{sec:QBP}

Consider now the quantum case $H = \sum_{\langle i,j\rangle}h_{i,j}$ where $h_{i,j}$ are Hermitian operators acting on site $i$ and $j$ of a chain. Defining the $\odot$-product between positive operators $A\odot B  = e^{\log(A) + \log(B)}$, we can write the partition function in a form very similar to \eq{Zclass} 
\begin{align}
\cZ(\beta) &= \Tr( e^{ -\beta H}) \\
& = \Tr (e^{-\beta h_{N,N-1}}\odot \ldots e^{-\beta h_{3,2}}\odot e^{-\beta h_{2,1}}) \label{eq:Zquantum}
\end{align}
with the sums replaced by traces and products by $\odot$-products. This expression cannot be reorganized like its classical counterpart because--unlike ordinary products--the $\odot$-product does not obey a distributive law in the sense that $\Tr_a(e^{-\beta h_{c,b}} \odot e^{-\beta h_{b,a}} ) \neq e^{-\beta h_{c,b}} \odot \Tr_a (e^{-\beta h_{b,a}} )$. 

However, the distributive law holds when the chain forms  a quantum Markov network \cite{LP08a}, i.e. $I(a:c|b) = 0$ where the quantity $I(a:c|b) = S(a,c) + S(b,c) - S(b) - S(a,b,c)$ is known as the conditional mutual information and $S$ is the von Neumann entropy. Intuitively, this condition means that all correlations between $a$ and $c$ are mediated through $b$. This does not prevent $a$ and $c$ from being correlated, but implies that all information learned about $a$ by measuring $c$ can equivalently be obtained by measuring $b$ instead. While this condition does not hold for generic quantum interactions, it becomes approximately true when the ``Markov shield" $b$ is sufficiently thick. 

To understand this behavior physically, consider again a one-dimensional chain  with nearest neighbor interactions $H = \sum_{i} h_{i,i+1}$. The thermal Gibbs state at inverse temperature $\beta$ is given by $\rho =e^{-\beta H}/\cZ(\beta)$. The reduced state of spins $2,3,\ldots ,N$ is obtained by tracing out the first spin of the chain, i.e. $\rho_{2,\ldots,N} = \Tr_{1}( \rho)$. We can formally define the {\em effective thermal Hamiltonian} $H_{\rm eff}$ acting on sites $2,\ldots, N$ up to normalization by 
\begin{equation}
\rho_{2,\ldots,N} = e^{-\beta H_{\rm eff}}.
\end{equation} 
In other words, $H_{\rm eff}$ is the traceless part of $-\frac 1\beta {\rm Log}( \rho_{2,\ldots N})$ where Log is the principal matrix logarithm. Similar effective Hamiltonians have also been studied in \cite{PE09a} in the context of reduced density matrices of fermionic and bosonic ground states.  We also define the {\em effective thermal potential} $V = H_{\rm eff} - \sum_{i=2}^{N-1} h_{i,i+1}$ as the term added to the bare Hamiltonian on sites 2 to $N$ due to the presence of site 1.  

\begin{figure}
\center \includegraphics[width=8cm]{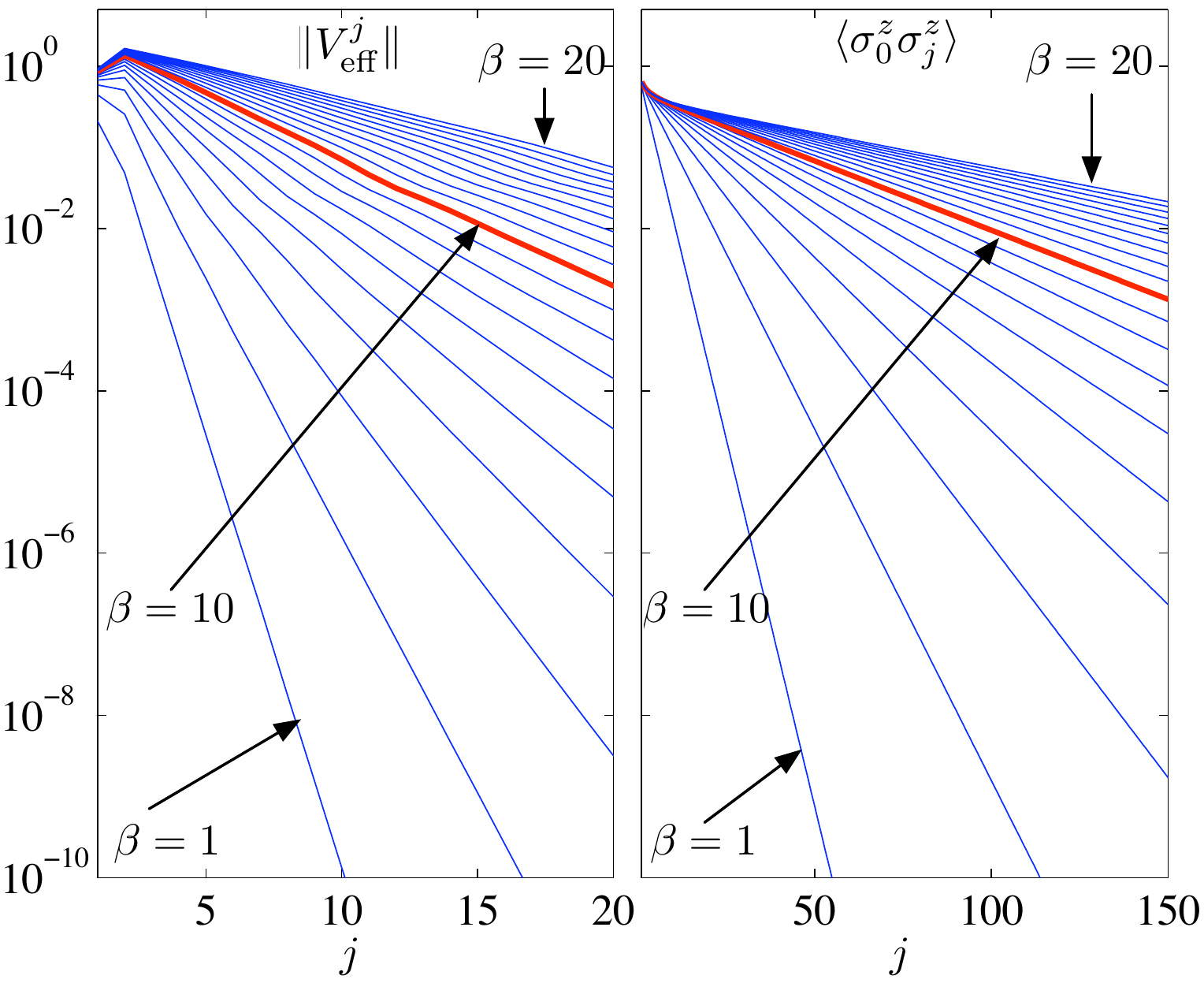} 
\caption{(Color online) Upper bound on $||V_{\rm eff}^j||$ ({\it left}) and correlations ({\it right}) of the Ising chain with critical transverse field. These exact values are obtained from a Jordan-Wigner transform. Note that $||V_\infty^j||$ decays much faster than the correlations (notice the different length-scales). The values for $\beta=10$ are in a different color as an aid to the eye. }
\label{Vplot}
\end{figure} 

At high temperatures, the effective thermal potential is short ranged. Consider for instance the one-dimensional Ising model with transverse field on an infinite chain $H = \sum_{i=-\infty}^\infty \sigma_i^z \sigma_{i+1}^z + B \sigma_i^x$. At zero temperature, this model exhibits a phase transition at the critical transverse field value $B = 1$. Fig.~\ref{Vplot} shows the value of the effective potential $V_{\rm eff} = -\frac{1}{\beta} {\rm Log} \Tr_{-\infty,\ldots,0} (e^{-\beta H}) - H_{\rm bare}$ obtained from cutting the critical Ising chain in half, i.e. tracing out spins $-\infty$ to 0 from the thermal state of an infinite chain. What is plotted is (an upper bound to) the operator norm of the cumulants of  $V_{\rm eff}$: $V_{\rm eff}^j = \Tr_{j+1,\ldots,\infty}( V_{\rm eff} - \sum_{k=1}^{j-1} V_{\rm eff}^k)$ with $V_{\rm eff}^1 = \Tr_{2,\ldots,\infty}(V_{\rm eff})$. We see that $V_{\rm eff}$ has a very short range,  in fact much shorter than the correlation length in the system.

\begin{figure}
\center\includegraphics[width=\columnwidth]{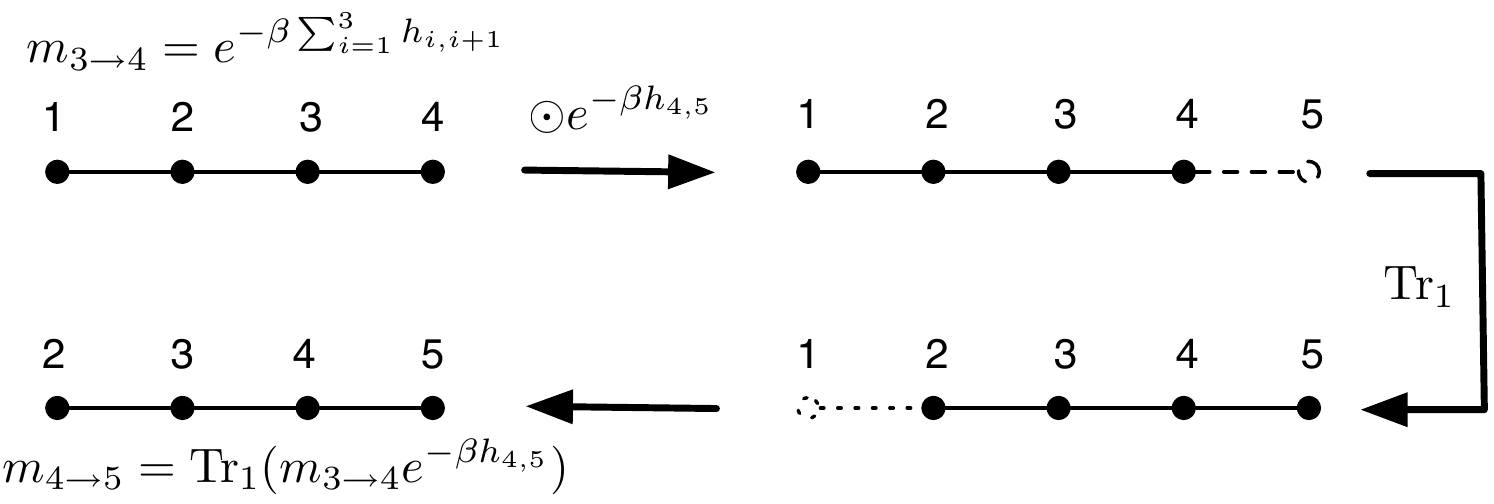}
\caption{Calculating $m_{4\rightarrow 5}$ from $m_{3 \rightarrow 4}$ in an iteration of BP algorithm with $l = 4$.  In the first step $e^{h_{4,5}}$ is added to $m_{3 \rightarrow 4}$ using the $\odot$-product.  Then, the first spin is traced out yielding $m_{4\rightarrow 5}$.}
\label{messagepassing}
\end{figure}

The distributive law becomes approximately true when the distance between the traced-out site and the end of the cluster is sufficiently large to support the effective thermal potential. Thus, we can reorganize the sum of \eq{Zquantum} as in the classical case, but keeping an $l$-site Markov blanket between the traced-out site and the end of the cluster:
\begin{widetext}
\begin{align}
\cZ(\beta) 
&= \Tr_{1,\ldots, N} \left( e^{-\beta h_{N-1,N}}\ldots \odot e^{-\beta h_{1,2}} \right) \\
&\approx \Tr_{2,\ldots, N}\left( e^{-\beta h_{N-1,N}}\ldots \odot e^{-\beta h_{l+1,l+2}} \odot \Tr_1\left( e^{-\beta h_{l,l+1}}\odot\ldots\odot e^{-\beta h_{1,2}} \right)\right) \\
&\approx \Tr_{3,\ldots, N}\left( e^{-\beta h_{N-1,N}}\ldots \odot e^{-\beta h_{l+2,l+3}} \odot  \Tr_2\left(e^{-\beta h_{l+1,l+2}} \odot m_{l\rightarrow l+1} \right)\right) \\
& \approx \ldots \\
& \approx   \Tr_{N-l,\ldots,N}\left(e^{-\beta h_{N-1,N}} \odot m_{N-1\rightarrow N} \right) 
\end{align}
\end{widetext}
where we have defined $m_{i\rightarrow i+1} = \Tr_{i-l+1}(e^{-\beta h_{i,i+1}}\odot m_{i-1\rightarrow i})$, which is an operator acting on sites $i-l+2$ to $i+1$ (see Fig. \ref{messagepassing}). These equations require manipulating operators on up to $l+1$ spins, so their complexity increases exponentially with the window size $l$. 

Like in the classical setting \eq{MP}, these message passing rules can be generalized to arbitrary graphs, enabling the computation of one- and two-body beliefs from which various quantities of interest such as energy can be computed. The method can also be adapted to estimate all correlation functions. 

Because of computational limitations, the effective thermal potential $V_{\rm eff}$ cannot be computed exactly. Instead, we can only estimate its value on a cluster of finite size. Thus, the main source of error in our method is due to the truncation of $V_{\rm eff}$. We can assess the error by evaluating the portion of $V_{\rm eff}$ we discard. On a chain for instance, the error caused on the estimate of the beliefs is
\begin{equation}
\frac 1{\cZ(\beta)} (e^{-\beta H_{\rm eff}} - e^{-\beta (H_{\rm eff} - \sum_{j>l} V_{\rm eff}^j)} ) \approx \beta \sum_{j>l} \langle V_{\rm eff}^j \rangle
\end{equation}  
for $\beta V_{\rm eff}^j \ll 1$. Making the assumption (see Fig. \ref{Vplot}) that $\|V_{\rm eff}^j\|$ decreases exponentially with $j$, we estimate this quantity by  
\begin{equation}
\beta \langle V_{\rm eff}^{l+1}\rangle \approx \beta \langle V_{\rm eff}^{l}\rangle \(  \frac{\langle V_{\rm eff}^{l}\rangle}{\langle V_{\rm eff}^{l-1}\rangle}\)
\label{eq:errorIsing}
\end{equation}
which is our final error estimate. With the same reasoning, similar estimates can be derived for the error associated with different observables. 

\begin{figure}
\center\includegraphics[width=9cm]{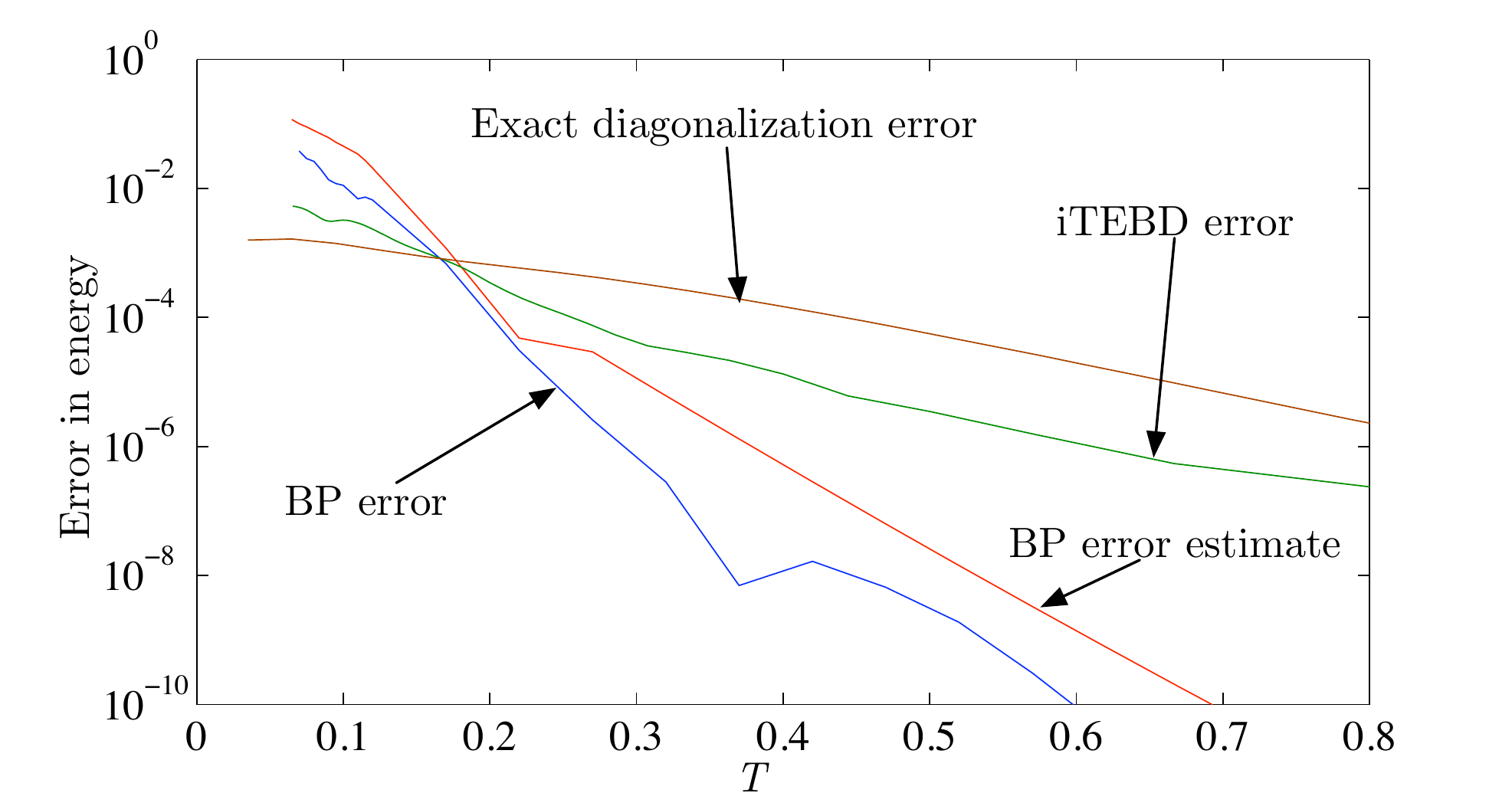}
\caption{Error on the energy density as a function of temperature for the critical Ising chain. For QBP with window size $l=10$, we show the error estimate of \eq{errorIsing} and the true error obtained by comparison with analytical solution.  Also shown is the finite size error for exact diagonalization of a 11-site chain and error caused by iTEDB \cite{Vid06a} with parameters $\delta T = 0.001$ and $\chi = 150$, both of which require equivalent computational resources.}
\label{errorboundfig}
\end{figure}

Fig. \ref{errorboundfig} compares this error estimate to the true error produced by BP for the evaluation of the energy density of the critical one-dimensional Ising chain. Clearly, the error estimate accurately bounds the true error. This figure also illustrates the power of BP by comparing the accuracy with which BP estimates the energy density of an infinite chain to what is achieved by brute force diagonalization with equivalent computational resources. BP largely outperforms diagonalization down to temperatures of order 0.1. This behavior is expected because, as seen on Fig. \ref{Vplot} (red line), the range of the effective thermal potential becomes larger than the window size ($l=10$) at this temperature. 

\section{Disordered system revisited}\label{sec:Cayley}

The use of belief propagation in physics originates in the study of disordered systems--spin glasses--where it is more often referred to as the ``cavity method'' \cite{MP01a}. Along with the sliding window algorithm outlined in the previous section, we presented in  \cite{LP08a,PB08a} a second distinct way to generalize BP to the quantum setting. This method, which we named ``replica BP'',  maps the quantum system to a classical system with one additional spatial dimension of length equal to the inverse temperature $\beta$. The edges of a graph become ribbons on which classical BP can be employed for sufficiently low $\beta$. This technique was independently introduced by Laumann {\it et al.} \cite{LSS07a} for the study of the transverse field quantum Ising spin glass.  The continuous imaginary-time limit of this procedure was later studied in \cite{KRSZ08a}.

The numerical results obtained in \cite{PB08a} suggest that, for a given amount of computational power, sliding window BP is much more accurate than replica BP. In this section, we revisit the spin-glass model of \cite{LSS07a} using sliding window BP, and apply the method outlined in the previous section to estimate the accuracy of our results.  

The spins are located at the vertices of a degree-3 Cayley tree. The Hamiltonian has Ising coupling between neighboring spins and a transverse field $B$:
\begin{equation}
H = \sum_{\langle ij\rangle} \sigma_i^z \sigma_j^z + B \sum_i \sigma_i^x + \sum_{i \in {\rm Boundary}} r_i \sigma_i^z.
\end{equation}
The last term is a random parallel boundary field introduced to create frustration in the system. The strength of the boundary fields $ r_i$ are chosen at random uniformly in $[-1,1]$. The quantity of interest in this setting is the Edwards-Anderson order parameter $q_{\rm EA} = \langle \sum_j \langle \sigma_j^z\rangle^2\rangle_Q$ on the lattice at the thermodynamic limit, where $\langle \cdot \rangle_Q$ refers to the quench average over the random boundary field configurations. Note that this order parameter is defined along the axis perpendicular to the applied external field $B$, but parallel to the random boundary field. Hence, $q_{\rm EA}$ is zero in the paramagnetic phase. It becomes non-zero on the onset of the glassy phase where the system settles into one of many meta-stable randomly polarized state.

\begin{figure*}
\center\includegraphics[width=12cm]{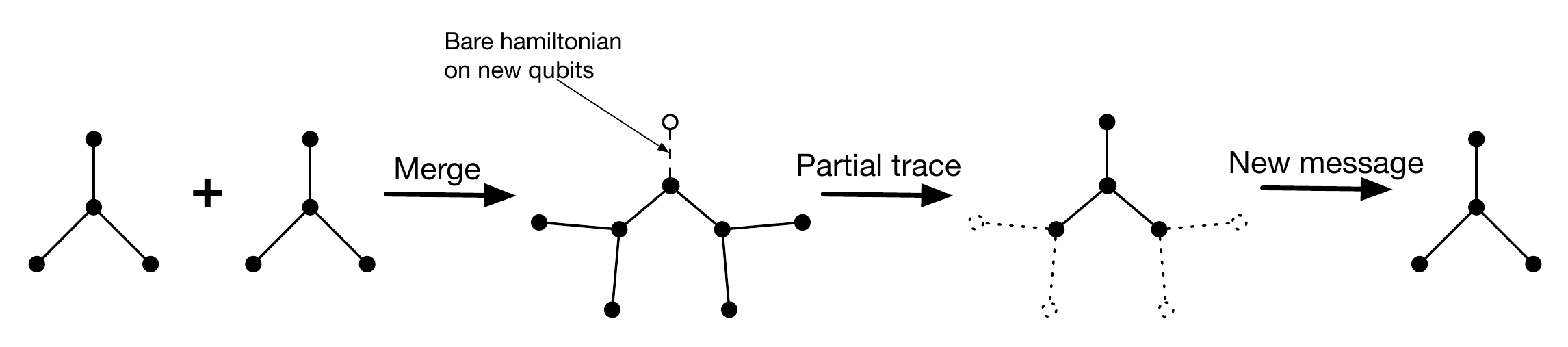}
\caption{Schematic illustration of the procedure to calculate the message from a node to its parent.  First, two messages from its children are merged using the $\odot$-product .  The bare hamiltonian term relating the current node to its parent is also added.  Tracing out the leaves of this 8-spin message gives the message to the node's parent.  }
\label{cayleymessage}
\end{figure*} 

Our numerical simulations are performed on a tree of depth 12. To further reduce the finite-size effects, the EA order parameter is only evaluated on the central spin of the lattice, away from the boundary. The order parameter can be computed from the single-body belief $b_j$ obtained from BP.   On a chain, we would start from one end  and propagate messages to the other end of the chain.  On the tree, we start from the leaves, and propagate messages towards the middle.  The basic message passing step is demonstrated in Fig. \ref{cayleymessage}.  First, two messages from the children of a given node are combined using the $\odot$-product .  The bare Hamiltonian term relating the node to its parent is then added.  Tracing out the leaves of this 8-spin message gives the message to the node's parent.  This procedure is repeated until the central site is reached where three messages are joined to produce the belief. The quench average is obtained by repeating this procedure 100 times with different boundary field configurations. 

\begin{figure}
\center\includegraphics[width=\columnwidth]{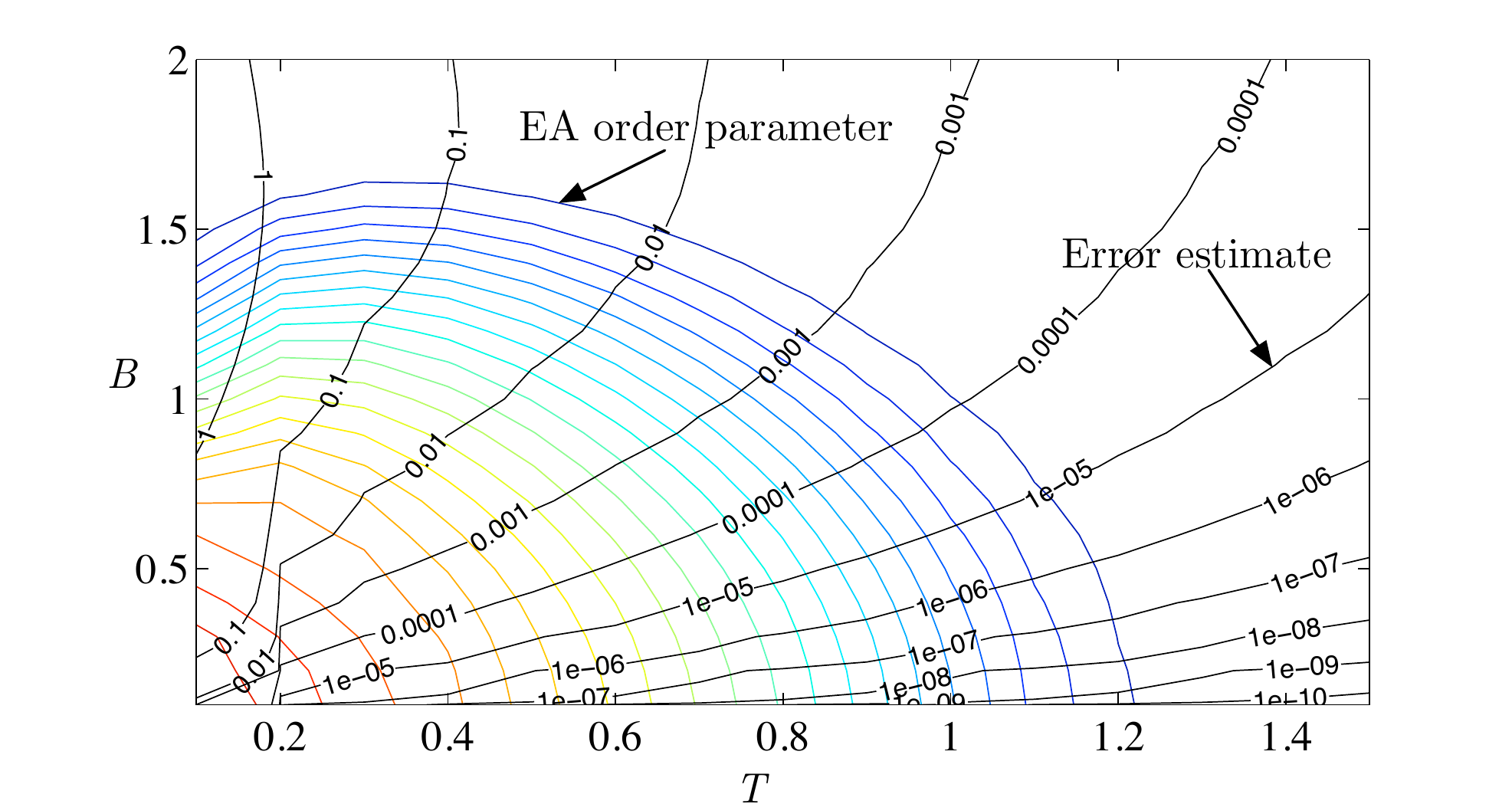}
\caption{The Edwards-Anderson order parameter for transverse field quantum Ising spin glass on a degree-3 Cayley tree with random parallel boundary fields.  The total depth of the tree is 12, and the plots show the average of 100 instances of random boundary fields.  Values of $q_{\rm EA}$ range from 0 to 1 and contours are equally spaced. The error on the order parameter is estimated using the procedure outlined at the end of Sec.~\ref{sec:QBP}. }
\label{cayleyresult}
\end{figure}

Because the graph contains no loops, the error estimate presented in the previous section is reliable.  However, note that the statistical fluctuations of the quench average are not included in this error estimate.  The statistical fluctuations of the average of $q_{EA}$ over the many instances of boundary fields range from 2 percent at low temperatures to 14 percent at high temperatures.  Therefore, the main source of error in parts of Fig. \ref{cayleyresult} is the statistical fluctuations which are not shown in the plot, and can be systematically reduced by increasing the sample size.

Fig \ref{cayleyresult} shows the EA order parameter $q_{\rm EA} $ in the transverse field-temperature diagram, along with the estimated BP error.  The glassy/paramagnetic phase transition line  agrees with results of \cite{LSS07a} up to the statistical fluctuations above $T\gtrsim0.3$.  At low temperatures ($T\lesssim 0.3$), our results indicate a phase transition line with a decreasing value of $B$ as $T$ is lowered. We could not think of any physical mechanism that could explain this behavior.  Moreover, this happens in a region of the phase diagram where the BP error is high. Hence, we suspect that the true phase transition line has a monotonous behavior in temperature and that the glassy phase persists all the way to zero temperature for a transverse field $B \lesssim 1.65$. This conclusion and, more generally, our entire phase diagram is in very good agreement with that of \cite{LSS07a}.

\section{Coarse grained belief propagation}\label{sec:CGBP}

We have seen that BP provides reliable estimates of thermal expectation values as long as the effective thermal potential $V_{\rm eff}$ is sufficiently short ranged to be tracked numerically. As seen e.g. in Fig. \ref{Vplot}, the range of $V_{\rm eff}$ grows linearly with $\beta$, so BP becomes unreliable at low temperatures. To probe lower temperatures, one needs to increase the window size $l$, which is not feasible because resources scale as $O(2^l)$.   On the other hand, as we lower the temperature, high energy excitations become increasingly irrelevant. This fact leads to efficient algorithms for zero temperature simulations such as entanglement renormalization and DMRG. These algorithms become rapidly inaccurate at finite temperature because they are only able to keep track of a small number of eigenstates. In this section, we will describe a method that interpolates between BP at  high $T$ to ER at $T=0$. Before we do so, we briefly review ER, see \cite{Vid05a,EV08a} for a detailed description.

\subsection{Entanglement renormalization}

\begin{figure}
\center\includegraphics[scale=0.25]{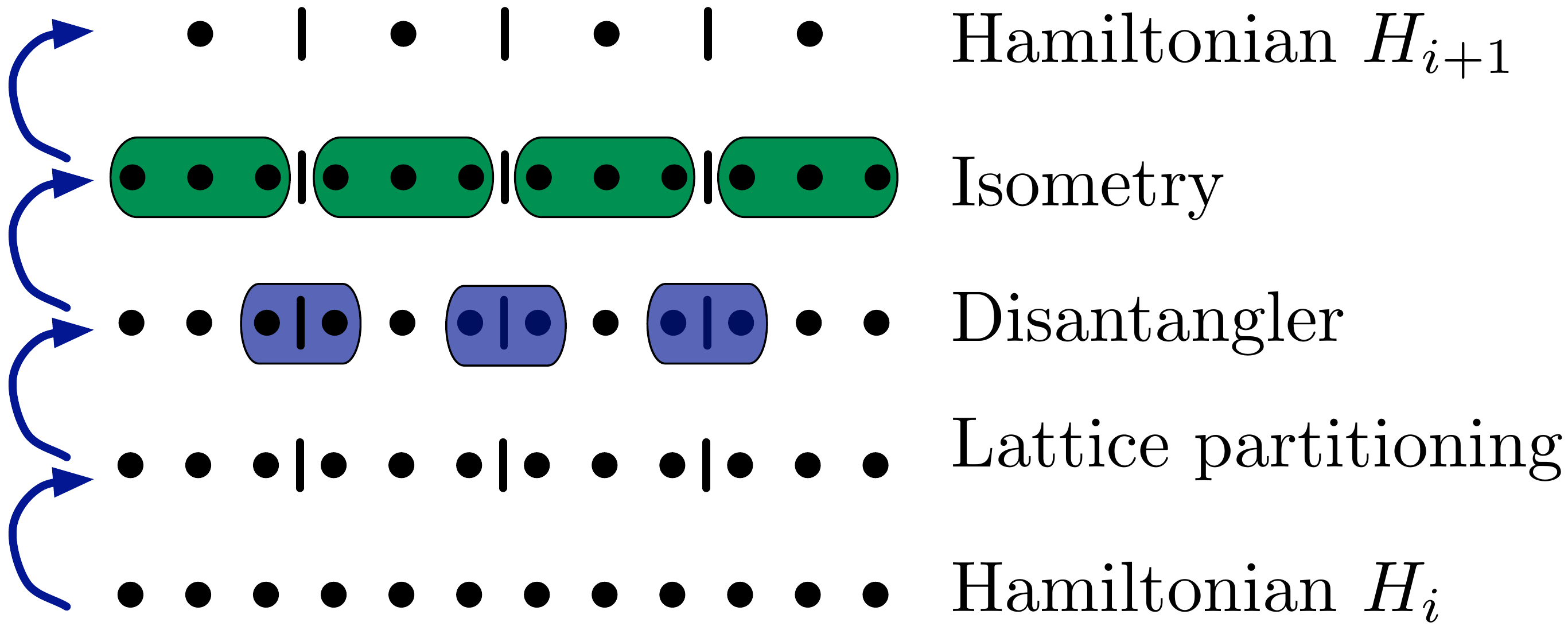}
\caption{Schematics of entanglement renormalization for a ternary MERA of a one dimensional lattice.}
\label{fig:ER}
\end{figure}

Entanglement renormalization \cite{Vid05a,EV08a} builds on the multi-scale renormalization ansatz (MERA) which asserts that certain degrees of freedom can be decoupled from the ground state of local Hamiltonians by unitary transformations acting on small spatial regions. A concrete example of this scheme is illustrated in Fig.~\ref{fig:ER} in the case of a one dimensional lattice. The lattice is first partitioned into clusters each containing 3 consecutive sites. A disentangling transformation $u$ [a unitary transformation on $(\mathbb{C}^\chi)^{\otimes 2}$] is applied on the boundary of each cluster in order to minimize the correlations between neighboring clusters. Finally, local degrees of freedom are discarded from each cluster by means of an isometry\footnote{At the first iteration of ER, the disentangler would be a unitary transformation on  $(\mathbb{C}^d)^{\otimes 2}$ and the isometry would map $(\mathbb{C}^d)^{\otimes 3}$ to $\mathbb{C}^d$ where $d$ is the number of levels of the particles forming the lattice.} $v$ mapping $(\mathbb{C}^\chi)^{\otimes 3}$ to $\mathbb{C}^\chi$.

This procedure is applied repeatedly. At each iteration, the disentanglers and isometries transform the Hamiltonian $H_i$ to a new Hamiltonian $H_{i+1}$ acting on a smaller lattice and retaining only the lowest eigenstates of the previous Hamiltonian. ER is halted when only a few sites remain in the lattice so it can be handled exactly numerically. The disentanglers and isometries are chosen in such a way that the final state (or more generally subspace) minimizes the energy of the initial Hamiltonian $H_0$. This minimization problem is in general hard, but good heuristics have been devised for it \cite{Vid05a,EV08a}.

\subsection{The interpolated scheme: CGBP}\label{subsec:CGBP}

Both ER and BP revolve around the idea that some correlations are short range in the state of interest. BP becomes exact when the conditional mutual information $I([-\infty,j] : [j+l,\infty] | [j+1,j+l-1] )$ vanishes for a sufficiently large window size $l$. In other words, BP can work in the presence of arbitrary long range classical correlations but the purely quantum correlations must be short ranged. These quantum correlations tend to increase like the inverse temperature $\beta$ (c.f. Fig. \ref{Vplot}), which limits BP to relatively high temperatures, unless the window size $l$ can somehow be increased while keeping computational cost low.  In contrast, ER makes use of disentanglers to eliminate short-range quantum correlations in the system and coarse grain the lattice. Because it only keeps a few  low-energy states, it is limited to very low temperatures.

\begin{figure}
\center\includegraphics[width=\columnwidth]{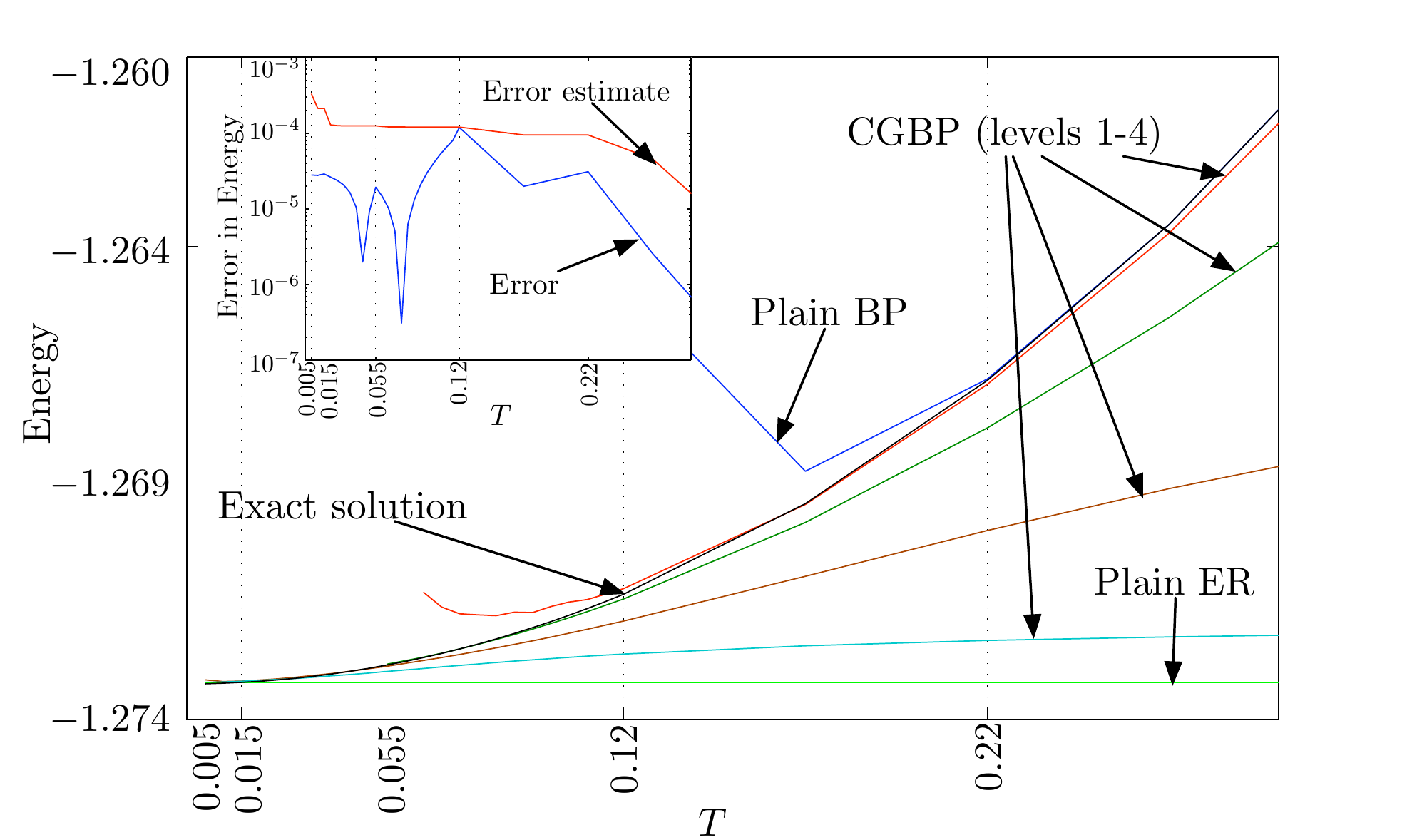}
\caption{(Color online) Energy {\it vs} temperature for the one dimensional critical quantum Ising model. Presented are the plain BP results for $l=10$ ({\it dark blue}), CGBP results for various levels of coarse graining with $\chi = 4$ and $l=5$ ({\it various colors}), the plain ER result $\chi = 4$ ({\it light green}) along with the exact result obtained from Jordan-Wigner transformations ({\it black}). The labels on the temperature axis correspond to the switching temperatures between various levels of coarse graining in the CGBP algorithm. E.g., plain BP is very accurate down to $T_1 \approx 0.22$, where BP combined with one level of coarse graining ({\it red}) becomes more accurate.  The inset shows the absolute error in the combined CGBP result with respect to the exact solution ({\it blue}), and the error estimate calculated using the procedure given in Sec. \ref{subsec:CGBP}.  While plain BP and plain ER are both very inaccurate for temperature range $0.005 \lesssim T \lesssim 0.22$, combining the two using CGBP yields very accurate results. }
\label{energykai4l5}
\end{figure}

Coarse grained belief propagation interpolates between these two methods and provides accurate thermal expectation values over a very large range of temperatures. At high temperature, CGBP reduces to ordinary BP. As the temperature is lowered, the error attributed to BP increases. At some temperature $T_1$, it becomes favorable to coarse grain the lattice using one step of ER. This procedure discards some high energy states, leading to a systematic error, but it effectively increases the BP window length $l$ by a constant factor (3 in the ternary ER scheme illustrated in Fig. \ref{fig:ER}). This increase in the window length improves the accuracy of BP and compensates for the loss of high energy states. As the temperature is lowered, the lattice is further coarse grained at $T_2$, $T_3$, ... until the CGBP reduces to ordinary ER. Fig. \ref{energykai4l5} illustrates this behavior for the one dimensional critical quantum Ising model.  Each coarse graining level provides a reliable estimate only for a small temperature range, but the union of these ranges cover the entire temperature domain. Thus, CGBP provides accurate estimates of thermodynamical observables in temperature ranges that are accessible to neither BP nor ER.

To calculate the optimal coarse-graining temperatures $T_i$, we would need independent error assessments for ER and BP. Then, we could switch to a coarser lattice whenever the increase in window size compensates for the coarse graining of the Hamiltonian. However, we are not aware of a reliable method to estimate the error caused by ER. Instead of using error estimates, we simultaneously perform BP on two different coarse grained levels and determine the switching temperatures by comparing the results. 

\begin{figure}
\center\includegraphics[width=\columnwidth]{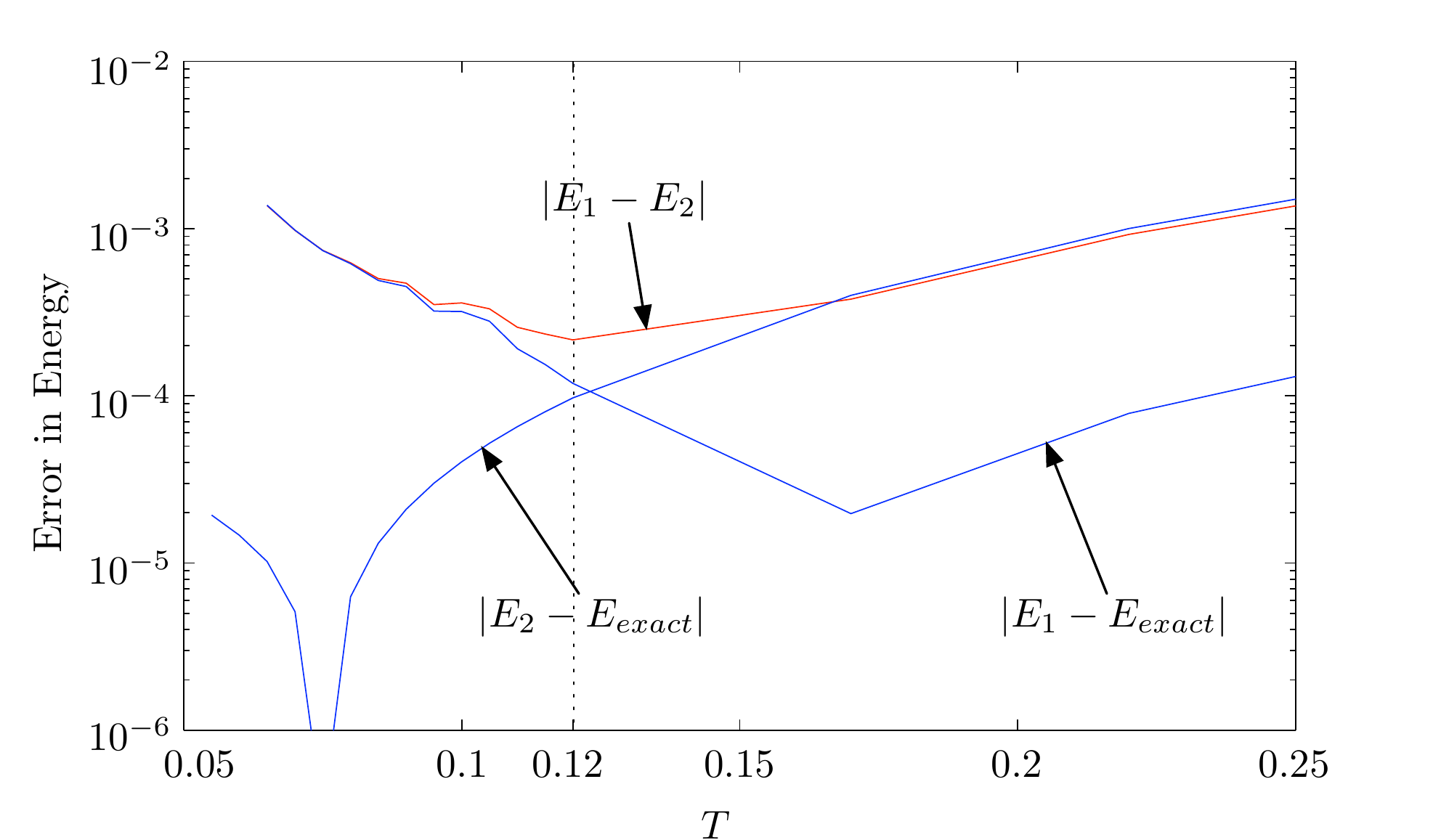}
\caption{The exact error in the energies calculated at the first and second levels of coarse graining, $|E_1-E_{exact}|$ and $|E_2-E_{exact}|$ are plotted along with $|E_1 - E_2|$.  Note that for both $|E_1-E_{exact}|$ and $|E_2-E_{exact}|$, MERA error dominates at high temperatures, and BP error dominates at low temperatures.  The minimum of their difference $|E_1 - E_2|$ occurs when the BP error at level 1 of coarse graining is equal to the MERA error at level 2 of coarse graining.  Furthermore, the high temperature portion of $|E_1 - E_2|$ is dominated by the MERA error at level 2 and the low temperature portion of $|E_1 - E_2|$ is dominated by the BP error at level 1.}
\label{diffE1E2}
\end{figure}

More precisely, let $\overline x_i(T)$ be the expectation value of some observable $X$ at temperatue $T$ obtained by BP on the $i^{th}$ level of coarse graining. To determine the switching temperature $T_{i+1}$  from the $i^{th}$ level of coarse graining  to the $(i+1)^{th}$, we calculate both $\overline x_i$ and $\overline x_{i+1}$ as we slowly lower the temperature. At high temperatures, the dominating error is attributed to the discarded high energy states, so $\overline x_{i}$ is more accurate than $\overline x_{i+1}$.  On the other hand, at low temperatures, as the range of the effective thermal potential gets larger than the BP window,  $\overline x_{i+1}$ becomes more accurate than $\overline x_{i}$. The two sources of errors are balanced when $|\overline x_i-\overline x_{i+1}|$ reaches a minimum (see Fig. \ref{diffE1E2}), so $T_i$ should be chosen at the position of this minimum. There can be exceptions to this rule that  result from accidental crossings of $\overline x_{i+1}$ and $\overline x_i$. In that case, the error estimate for BP presented in Sec. \ref{sec:QBP} can be used to discriminate between the multiple minima. Indeed, the switching should occur when the value of  $|\overline x_i-\overline x_{i+1}|$ is the closest to the error attributed to BP on the $i^{th}$ level because both numbers are estimates of the BP error.  The switching temperatures for Fig.~\ref{energykai4l5} were chosen following this method.

We can use the same reasoning to estimate the total error $\delta x(t)$ on our final estimate $\overline x(T)$ obtained by joining the $\overline x_i(T)$ over their respective range. We define $\delta  x_i^{\rm BP}(T)$ to be the error attributed to BP on the $i^{th}$ level of coarse graining. This quantity can be estimated as described in Sec. \ref{sec:QBP}. For $T > T_1$, we have $\delta x(T) = \delta  x_0^{\rm BP}(T)$ since BP is the only source of error. Between $T_1$ and $T_2$, there are two contributions to the error: the error $\delta  x_1^{\rm BP}(T)$ attributed to BP on the $1^{st}$ coarse grained level and the error attributed to ER caused by discarding high energy states. This second error decreases as temperature is lowered, and at $T = T_1$ it is equal to the BP error (this is how $T_1$ was defined). Thus, we obtain for $T_2\leq T < T_1$ the bound $\delta x(T) \leq \delta x_0^{\rm BP}(T_1) + \delta  x_1^{\rm BP}(T)$. More generally, for $T_{i+1}\leq T < T_i$ we get $\delta x(T) \leq \sum_{j \leq i}\delta x^{\rm BP} _{j-1}(T_{j})+ \delta  x_i^{\rm BP}(T)$.  See the inset of Fig. \ref{energykai4l5} for the error estimate of CGBP with $\chi = 4$ and $l=5$.

In addition to CGBP with $\chi = 4$ and $l = 5$, we have also investigated various other combinations of $\chi$ and $l$.  At equal computational costs---which are of $O(\chi^{3l})$---the results are qualitatively equivalent in the sense that they have roughly equivalent maximum error. However, varying $\chi$ and $l$ at fixed computational cost can improve the results for a given temperature. For instance, at very low temperatures, the computations with a higher $\chi$ yield results with better accuracy. On the other hand, larger $l$ and smaller $\chi$ perform better at high temperatures. Thus, the values of $\chi$ and $l$ could also be varied dynamically in the simulation, but we leave out this possibility for the moment.

The results we obtain with CGBP compare favorably with results obtained by other methods using equivalent computational resources. This can be seen by comparing the inset of Fig.~ \ref{energykai4l5} to the various curves shown on Fig.~\ref{errorboundfig}. The CGBP result with $\chi=4$, $l=5$ has at least four digits of accuracy for the entire temperature range.  The complexity of this simulation is equivalent to exact diagonalization of a 11-site chain, that produces a result accurate to roughly three digits. Many methods rely on a Trotter-Suzuki decomposition by discretizing imaginary time in intervals $\delta T$, which creates a systematic bias $O(\delta T^3)$ in the result. This is the case for instance of imaginary time dependent block decimation \cite{Vid06a}. This algorithm has complexity $(\chi d^2)^3 T_{\rm min}/\delta T$ where  $d=2$ for the Ising model is the dimension of the spins. The parameters $\chi = 150$ and $\delta T = 0.001$ yield the same complexity as the $\chi=4$ $l=5$ CGBP algorithm, and produce an error of roughly $10^{-2}$ at sufficiently low temperatures.

\section{Discussion}

The concept of an effective thermal potential, obtained by tracing out sites from the thermal Gibbs state, gives a clear physical picture of the workings of belief propagation. Adding a site to a spin chain can modify the thermal state even far away from the added site, as far as the correlation length of the system. However, this effect can be mimicked by adding a short range thermal potential to the Hamiltonian of the original spin chain. At finite temperature, we have seen that the range of this thermal potential can be much shorter than the correlation length of the system, making it more suitable for numerical simulations. This underlies the success of belief propagation and provides a means to assess its accuracy. We have illustrated these concepts and methods on the critical quantum Ising chain and the transverse field quantum Ising spin glass. For this last system, our findings are, within estimated error bars, in agreement with those of \cite{LSS07a} obtained from a different belief propagation implementation.  

At lower temperatures however, the range of the effective thermal potential becomes too large to handle numerically. For these temperature ranges, we have introduced the  coarse grained belief propagation algorithm by combining belief propagation with entanglement renormalization. Coarse graining discards some high-energy states, which leads to a systematic error in the thermal states.  On the other hand, it increases the accuracy of BP by shortening the range of the effective thermal potential by a constant factor. CGBP seeks an optimal compromise between these two effects in order to accurately probe temperature regimes where neither ER nor BP are reliable. Thus, CGBP truly extends the domain of applicability of the two underlying approaches. Moreover, results obtained by CGBP compare favorably to other methods using equivalent computational resources.

The drawback of CGBP is that it inherits some intrinsic limitations of the underlying approaches. For instance, ER is applicable only when there exists  a coarse graining method which preserves the locality of the Hamiltonian.  For graphs with exponential spreading structure, such as the Cayley tree studied in Sec. \ref{sec:Cayley}, we are not aware of such coarse graining procedures. This is the reason we have not implemented CGBP on that system. Another limitation comes from the shortcoming of BP on graphs containing many small loops such as two-dimensional lattices. Classically, this limitation can be alleviated using generalized BP \cite{YFW02a,WJW03a}. We are currently working on combining such generalizations with ER to study lattices of higher dimension.

\medskip
\section{Acknowledgements}

Computational resources were provided by the R\'eseau qu\'eb\'ecois de calcul de haute performance (RQCHP). DP receives financial support from Canada's NSERC and le Fonds qu\'eb\'ecois de la recherche sur la nature et les technologies. EB is supported by DoE under Grant No. DE-FG03-92-ER40701, and NSF under Grant No. PHY-0803371.

\end{document}